# 2D Topological Edge States in Periodic Space-Time Interfaces

Ohad Segal[1], Yonatan Plotnik[2], Eran Lustig[3], Yonatan Sharabi[1], Moshe-Ishay Cohen[1], Alexander Dikopoltsev[4], and Mordechai Segev[1,2]

1. Electrical & Computer Engineering Department, Technion, 32000 Haifa, Israel
2. Physics Department, Technion, 32000 Haifa, Israel
3. Ginzton Laboratory and Dept. of Electrical Engineering, Stanford University, CA 94305
4. Institute of Quantum Electronics, ETH Zürich, Zurich, Switzerland

msegev@technion.ac.il

## Abstract

Topological edge states in systems of two (or more) dimensions offer scattering-free transport, exhibiting robustness to inhomogeneities and disorder. In a different domain, time-modulated systems, such as photonic time crystals (PTCs), offer non-resonant amplification drawing energy from the modulation. Combining these concepts, we explore topological systems that vary periodically in both time and space, manifesting the best of both worlds. We present topological phases and topological edge states in photonic space-time crystals – materials in which the refractive index varies periodically in both space and time, displaying bandgaps in both frequency and momentum. The topological nature of this system leads to topological invariants that govern the phase between refracted and reflected waves generated from both the spatial and the temporal interfaces. The 2D nature of this system leads to propagating edge states, and a unique edge state that grows exponentially in power whilst following the space-time edge.

In recent years, topological phases of matter have received great attention in many scientific communities. They were originally used to classify the structure of electronic bands of materials [1], but are now recognized as a universal wave phenomenon in a variety of physical systems. Topological phases were observed in microwave systems [2], photonics [3], cold-atoms [4], [5], acoustics [6], [7], exciton–polaritons [8], [9] and many more physical systems. The most intriguing and important implication of topological phases is the existence of topologically protected edge states which facilitate scattering-free transport of waves in a fashion that is robust to disorder and defects. This robustness lies at the heart of macroscopic quantum phenomena such as the quantum Hall effect [10] and exciting new materials such as topological insulators [4], [5], [11], [12], [13], [14], [15]. The topologically protected transport of waves appears in systems of two dimensions or more. In photonics, this was demonstrated by using judiciously designed platforms that produce topological edge states and robust propagation of light [2], [16], [17] and led to the development of devices such as topological insulator lasers [18], [19], [20], topological sources of quantum light [21] and topological protection of entanglement [22], [23].

More recently, topological phases were also found in 1D time-varying systems such as photonic time crystals (PTCs) [24]. A PTC is a homogenous medium whose refractive index is modulated periodically in time, with modulation contrast on the order of unity and modulation period on the order of a single optical cycle [25], [26]. Such strong modulation in time causes time-reflections (yet to be observed for optical waves) [27], [28], [29] and time-refractions [30], [31], [32], which interfere, giving rise to a band-structure with bands separated by gaps in the momentum. The electromagnetic modes associated with the momentum band-gaps exhibit exponential increase (or decrease) in their amplitude, extracting energy from the modulation (or transferring energy to it) and were observed at RF frequencies [26] and very recently in microwaves [33]. Because of this exponential increase to their amplitude, the modes residing in the momentum gaps of PTCs have been suggested as a source of radiation and non-resonant amplification of light [34], [35], not relying on any atomic resonance and not requiring population inversion.

Several years ago, our group predicted topological phases in PTCs [24], which were subsequently demonstrated in a synthetic-space system [36]. However, since time is one-dimensional (at least to the best of our current knowledge), the topological edge states found in PTCs are fundamentally zero-dimensional, and hence cannot support topologically-protected transport along the edges of the system. This understanding raises a natural question: is it possible to add a spatial dimension with periodicity, create a space-time photonic crystal [37], [38], and design a 2D topological system displaying topologically-protected edge transport?

Here, we do exactly that: we present topological phases and topological interfaces in Photonic Space-Time Crystals (PSTCs), and discover edge states displaying topologically-protected transport. We design topological PSTCs exhibiting nontrivial topology by periodically modulating the permittivity of a medium both in space and in time. We find that this system displays a hybrid frequency-momentum gap governed by four topological invariants, predict the formation of topological edge states featuring robust transport, and simulate their evolution numerically. Intriguingly, we find that PSTCs display a new kind of topological edge state that accumulates power while moving on the space-time edge. This discovery opens up new possibilities for the manipulation and control of light, which could have far-reaching consequences in a variety of fields.

Photonic space-time crystals (PSTC) are a new class of materials whose electromagnetic properties (e.g., refractive index) are modulated periodically both in space and time [37], [38]. They exhibit bandgaps in both momentum and energy. This dual periodic modulation, with spatial and temporal periodicity at the scale of the wavelength and the optical cycle, endows PSTCs with unique properties. Unlike spatial photonic crystals, in which energy is conserved, or photonic time-crystals, in which momentum is conserved, in PSTCs generally both energy and momentum are not conserved. This lack of stringent conservation laws can give rise to complex wave dynamics where new frequencies and new momenta emerge, and the amplitudes of the waves might vary erratically, exhibiting huge variations in their amplitudes. At the same time, the lack of stringent conservation laws may also open up new research avenues and technological opportunities. For these reasons, finding topological phenomena and topological transport – which generally require strict conservation laws - is so surprising and fundamentally interesting in PSTCs.

As an instructive example, consider a dielectric medium which varies in both space and time, in which the permittivity varies as

$$\varepsilon(x,t) = \underbrace{\left\{1.5 + a_1 \cos\left(\frac{2\pi x}{\Lambda} + \phi_{x,1}\right)\right\}}_{\varepsilon_x} \underbrace{\left\{3 - b_1 \cos\left(\frac{2\pi t}{T} + \phi_{t,1}\right) - c_1 \cos\left[2\left(\frac{2\pi t}{T} + \phi_{t,1}\right)\right]\right\}}_{\varepsilon_t} \quad (1)$$

where $\Lambda$ and T are the periodicity in space and in time of the PSTCs, respectively. The constants are chosen such that $\varepsilon(x,t) \geq 1$ for all $x, t$. For a unit-cell centered in $(x,t) = (0,0)$ in space-time, the phases $\phi_{x,1}, \phi_{t,1} \in \{0, \pi\}$ dictate the space and time topological invariants, respectively. In Eq. (1) the parameters $a_1$ control the size of the energy

(frequency) gaps and $b_1, c_1$ control the size of the momentum (wavenumber) gaps in the PSTCs band structures. The ratio $R = cT/\Lambda$ controls the relative position between the energy and momentum gaps in the PSTCs band structure. As pointed out by [37], the PSTCs band structures may include momentum bandgaps, energy bandgaps and hybrid bandgaps. Importantly, the main features of the modes in the hybrid gaps are dictated by the 1D modes (space or time) that have a larger imaginary part in their eigenvalue [37].

We begin by comparing the topological phases of these PSTCs with the topological phases of the photonic space crystals $\varepsilon_x$ and photonic time crystals $\varepsilon_t$. The Zak phase of the photonic space crystal, $\varepsilon_x$ [39], [40], (which can be 0 or $\pi$) is a topological quantity calculated for each band below an energy gap; we refer to it as the "space topological invariant". Using this Zak phase, the topological invariant (also 0 or $\pi$) of each bandgap can be calculated by summing over the Zak phases of all the bands below the gap. The space topological invariant dictates the phase difference between an incident wave and a reflected wave from a spatial interface between $\varepsilon_x$ and a stationary homogeneous medium [40]. This space topological invariant also predicts the formation of a zero-dimensional edge state generated on a space interface between two crystals with opposite space topological invariant. In a similar manner, the time topological invariants of the photonic time crystal $\varepsilon_t$ is calculated for each band below a momentum gap [24]. Notice that, for the photonic time-crystal, the time topological invariants dictate the phase difference between the forward-propagating amplified waves and backward-propagating amplified waves, when an "incident" wave in a stationary homogenous medium starts experiencing the PTC. This time topological invariant also predicts the formation of a zero-dimensional edge state generated on a **time** interface between two photonic time crystals with opposite time topological invariant [24]. In the PSTCs of Eq.1, when considering a space interface, we find that the space topological invariants of the energy gaps of $\varepsilon(x,t)$ can be obtained solely from $\varepsilon_x$. Similarly, when considering a time interface, the time topological invariants of the momentum bandgaps of the PSTCs are obtained solely from $\varepsilon_t$. This concept was studied by our group [41] and was recently demonstrated in a synthetic space-time crystal [42]. To avoid confusion, henceforth we use 0 or $\pi$ to describe the space or time topological invariant and not bring up the Zak phase anymore.

Next, consider a generic interface between two different PSTCs, which, for simplicity, we sketch as a straight diagonal line in Fig. 1a. This system consists of two different PSTCs denoted by 1 and 2 with each of them occupying a half infinite space-time region. The two space-time crystals are separated by a space-time interface. The interface is smooth to maintain the general geometric features of the space-time crystal. We find that a topological edge state may form at this interface, depending on the topology of the bands. If the space-time topological invariants of PSTC 1 are different from those of PSTC 2, a topological edge state appears at the interface (as illustrated in Fig. 1c). Surprisingly, even though the permittivity is separable in $x$ and $t$, the topology allows for the edge state to form on generic interfaces that can cross the $x$-$t$ plane diagonally.

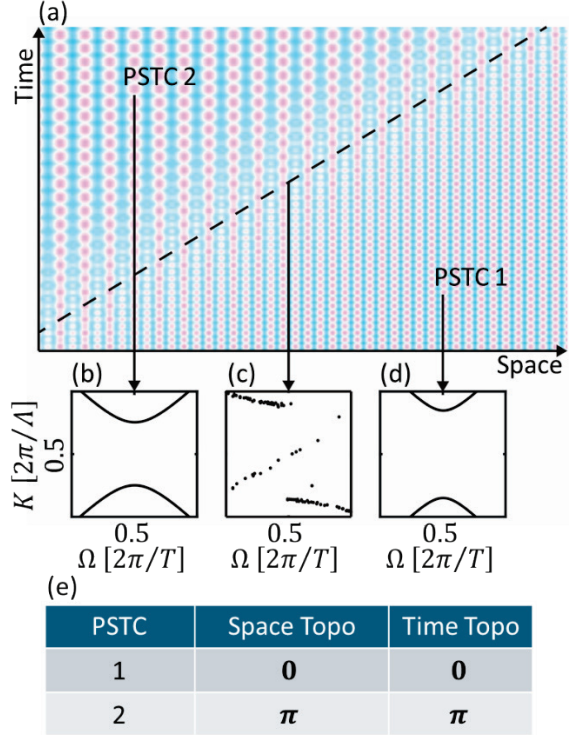

Figure 1: **Space-time interface (marked with a dashed line) formed between two space-time crystals**. (a) Example of two PSTCs and the space-time interface formed between them. (b) Band structure of PSTC 2. (c) Dispersion relation of the topological edge modes (central line) at the interface between PSTCs 1 and 2. These modes exist if the topological phase difference between the two PSTCs is nontrivial. The upper and lower branches are part of the space-time bulk. Notice that only one branch of topological edge modes is present, which implies unidirectional propagation. (d) Band structure of PSTC 1. (e) The space and time topological invariants of PSTCs 1 and 2 summarized in a table.

As a study case, we set the permittivity of PSTCs 1 and 2 to the permittivity function presented in Eq. 1, but differing from each other in their periodicities and phases:

$$\varepsilon_1(x,t) = \left\{1.5 + a_1 \cos\left(\frac{2\pi x}{\Lambda} + \phi_{x,1}\right)\right\}\left\{3 - b_1 \cos\left(\frac{2\pi t}{T} + \phi_{t,1}\right) - c_1 \cos\left[2\left(\frac{2\pi t}{T} + \phi_{t,1}\right)\right]\right\}$$
$$\varepsilon_2(x,t) = \left\{1.5 + a_2 \cos\left(\frac{2\pi x}{2\Lambda} + \phi_{x,2}\right)\right\}\left\{3 - b_2 \cos\left(\frac{2\pi t}{2T} + \phi_{t,2}\right) - c_2 \cos\left[2\left(\frac{2\pi t}{2T} + \phi_{t,2}\right)\right]\right\}$$
(2)

where $\Lambda$ and $T$ are the periodicity in space and in time of PSTC 1, respectively, and they differ from those of PSTC 2 by factor 2. The parameters $a_1, a_2, b_1, b_2, c_1, c_2$ and the ratio $R = cT/\Lambda$ are adjusted such that the band structure of both crystals displays only hybrid energy-momentum gap in which the energy gap is larger than the momentum gap. Thus, as pointed out by [37], the space-time band structure of both crystals exhibits only energy gaps and does not support growing bulk modes (See [43] section 1 for more details and chosen parameters). For each PSTC we calculate the space and time topological invariants separately. Figure 2 shows the space-time band structures and the topological invariants for different choices of $\phi_{x,1}, \phi_{x,2}, \phi_{t,1}, \phi_{t,2} \in \{0, \pi\}$.

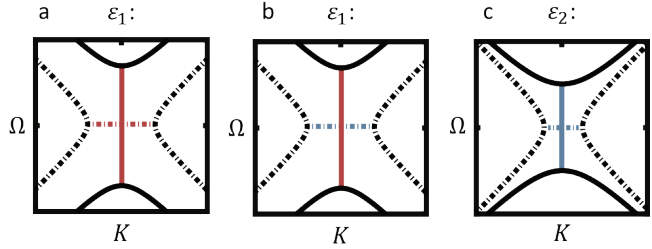

Figure 2: **Space-time band structure and topological phase of PSTCs 1 and 2 for different choices of $\phi_{x,1/2}, \phi_{t,1/2}$, zoomed on a specific bandgap**. Solid black lines correspond to the space-time band structure. Black dashed lines correspond to the underlying momentum gap of the time part of the PSTC, hidden by the bigger energy gap in the combined space-time band structure. The solid vertical line corresponds to an energy gap, and its color, red or blue, indicates that the space topological invariant is 0 or $\pi$. Likewise, the dashed horizontal line corresponds to a momentum gap, and its color, red or blue, indicates that the time topological invariant is 0 or $\pi$. **a)** $\phi_{x,1} = 0, \phi_{t,1} = \pi$. **b)** $\phi_{x,1} = 0, \phi_{t,1} = 0$. **c)** $\phi_{x,2} = 0, \phi_{t,2} = 0$.

The space invariant and time invariant can be 0 or $\pi$ in each PSTC, which yields four possible combinations. Each of these combinations exhibits different features, hence we discuss them separately:

I. The space invariant is the same in both PSTCs (space-trivial) and the time invariant is also the same in both PSTCs (time-trivial).

II. The space invariant is different between the two PSTCs (space-topological) while the time-invariant is the same in both PSTCs (time-trivial).

III. The space-invariant is the same in both PSTCs (space-trivial) while the time-invariant is different between the two PSTCs (time-topological).

IV. The space invariant is different between the two PSTCs (space-topological) and also time-invariant is different between the two PSTCs (time-topological). This last combination corresponds to the case presented in the table in Fig. 1e and to the resulting unidirectional topological edge state modes presented in Fig. 1c.

For each of the four combinations, we calculate analytically the interface modes and perform FDTD simulations (See [43] section 2 and 3 for calculations and FDTD details). Importantly, edge states only appear for combinations II and IV. The corresponding modal wavefunctions of these edge states are presented in Fig. 3. The edge states of combination IV have real Bloch wavenumber and Flouqet frequency (Fig. 3a,b) and therefore have wavefunctions that oscillate periodically in space and time. Their dispersion relation is similar to a light line, and hence they move with velocities similar to each other. As shown in Fig. 3c, each one of the modes is extended on the space-time edge while also behaving as a pulse with finite space and time domains (in fact, as explained in [43], there are two edges). The edge states differ in their wavenumber and frequency, their space-time shape, and their distance to the space-time edge. Some edge states are localized at the edge itself and some are localized a couple of lattice sites away from it. The edge states of combination II are fundamentally different: they exhibit real Bloch wavenumber and complex Flouqet frequency (Fig. 3d,e). These edge states have non-zero propagation velocity and accumulate power while moving on the space-time edge (Fig. 3f).

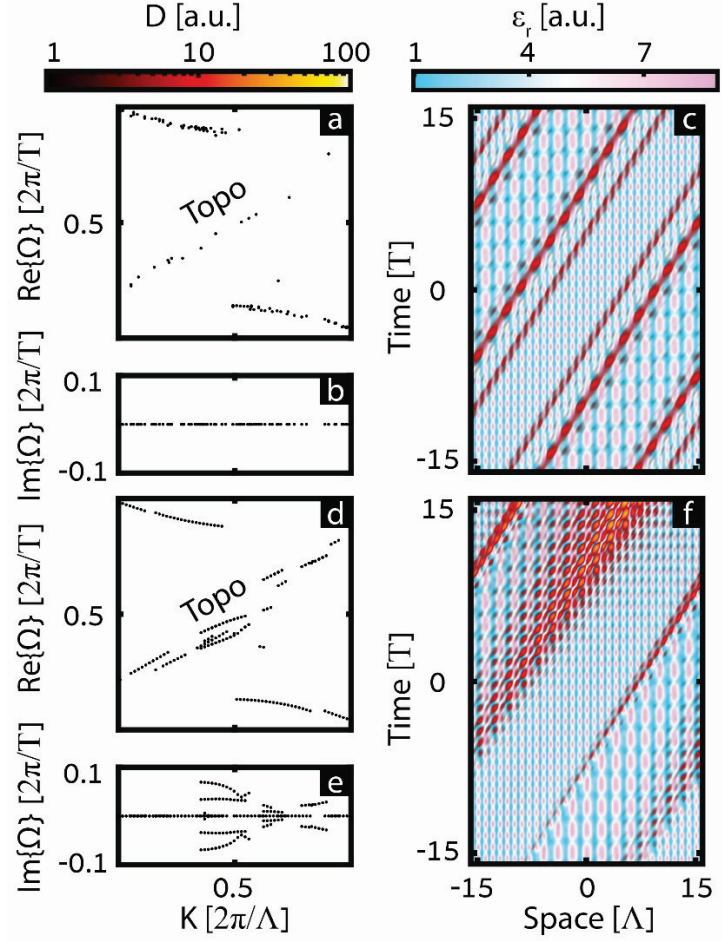

Figure 3: **Space-time edge states modes and band structures for combinations II and IV of the topological invariants**. (**a,b**) Real and imaginary parts, respectively, of the band structure of combination IV calculated using a superlattice with an interface between the two PSTCs. In combination IV, the imaginary part of the Floquet frequency is zero for all modes. (**c**) Two color maps overlay on top of each other: cyan-white-magenta marks the space time superlattice permittivity of combination IV and the black-red-yellow marks the intensity of the mode corresponding to the eigenvalue $(K, \Re\{\Omega\}) = (0.5, 0.5)$ from panel a). The displayed permittivity is a single unit cell of a superlattice. This unit cell contains finite bulks of PSTC 1 and PSTC 2 and the interfaces between them. (**d,e**) Real and imaginary parts of the band structure combination II. The topological modes have positive or negative imaginary Flouqet frequency which causes exponential increase (or decrease) of the field amplitude. (**f**) A growing topological edge state mode of combination II corresponding to the eigenvalue $(K, \Re\{\Omega\}) = (0.5, 0.5)$ from panel d).

An important property of topological edge states – in all physical systems - is their robustness to defects and disorder. Here, the edges are space-time, hence scattering from disorder in such structures is prone to exhibit new features, because (as far as we know) time is unidirectional. We explore the robustness of the edge states to disorder via FDTD simulations, where we introduce random fluctuations in the periodic space-time crystalline structure. Figure 4 shows an example of the simulated evolution dynamics of the space-time edge states for combinations II and IV of the topological invariants (for completeness, respective simulations for combinations I and III, which have no edge states, are presented in [43] section 3). The spectrum of the random variations used in all simulations are at frequencies ≤ than the frequencies defining the periodic time and space (crystalline) structure, and are of amplitude of ~1 (see [43] section 3 Eq. S9). As shown in Fig. 4b,d, the space-time fluctuations cause distortions in the periodic space-time structure, but the edge states remain unharmed, without any backscattering (since nothing can go back in time). Surprisingly, as Fig. 4a shows, in this combination the space-time fluctuations do not cause any noticeable time-reflection that would have scattered into the bulk of the space-time lattice. This is surprising since these fluctuations are essentially noise, and noise does not conserve energy nor momentum and hence may scatter waves out of the bandgap and into the bulk. ***Altogether, the transport of the edge mode shown in Fig. 4 displays the robustness characteristic of a topological edge state***. Likewise, all other edge modes of combination IV. On the other hand, for combination II (Fig. 4c), the fluctuations cause some time-reflection of the exponential tails of the amplified edge state.

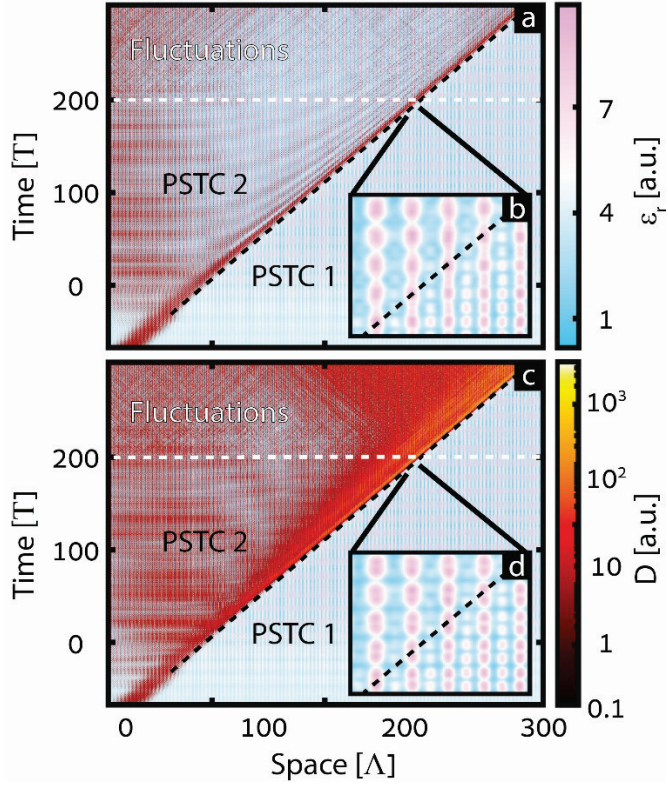

Figure 4: **FDTD simulated evolution dynamics of the space-time edge states for combinations II and IV of the topological invariants**. Two color maps are overlaid on top of each other, cyan-white-magenta colormap is the space time permittivity and black-red-yellow colormap is the electric field of the mode. The simulations study the evolution of a pulse directed at the space-time edge. The pulse enters the space-time permittivity from a stationary spatially-homogeneous permittivity through a gradual time-edge (shown in the figure at negative times). The space-time modulation begins at t=0. Only part of the pulse energy is transferred to the edge state modes and the rest is trapped in PSTC 2. After 200 temporal periods, fluctuations with frequencies as high as the time and space frequencies of the PSTCs are introduced. **a)** Evolution of a topological edge state packet of combination VI. The fluctuations do not cause back reflections or scattering into bulk modes. **b)** Zoom in on the space-time edge after the fluctuations begin. Random variation such as crooked lattice sites, can be observed on both side of the interface and on the interface itself. **c)** Evolution dynamics of a growing topological edge state of combination II. **d)** Zoom in on the space-time edge after the fluctuations start.

We emphasize that space-time topological edge states are fundamentally different from other topological systems as the space-time crystals do not conserve energy or momentum. Surprisingly, this does not hinder their robustness to disorder and defects.

The space-time topological edge states presented here offer several new features to the topological systems toolbox. First, they can grow (or decay) as they propagate along the edge, drawing power from the space-time modulation (or transferring energy to it). Second, these modes are spatio-temporal wavepackets, which means that turning on the space-time lattices generates ultra-short pulses from longer pulses or CW waves launched into the system. Last, from a fundamental science perspective, space-time edge states are a gateway to higher dimensional topological systems with new dynamics.

Finally, we note that experiments with space-time photonic crystals and their topological edge states are expected in the near future, most likely first in microwaves in systems similar to those employed in [29], [33] and subsequently in the optical regime, where recent advances with transparent conductive oxides hold the promise of observing PTCs [31], [32], [44], [45]. In addition, topological space-time PTCs can also be explored in experiments in synthetic dimensions, where recent experiments show much progress [42], [46], [47], and in metasurface platforms, where some of the features can already be observed [48].

This project was funded by grants from the US Air Force Office of Scientific Research (AFOSR) and the Breakthrough Program (Mapats) of the Israel Science Foundation.